\documentclass[11pt]{article}

\usepackage[final]{acl}

\usepackage{times}
\usepackage{latexsym}

\usepackage[T1]{fontenc}
\usepackage[utf8]{inputenc}

\usepackage{microtype}

\usepackage{inconsolata}
\usepackage{xurl}
\usepackage{graphicx}
\usepackage{booktabs}   % \toprule, \midrule, \bottomrule
\usepackage{tabularx}
\usepackage{array}
\usepackage{enumitem} 
\usepackage{array}
\usepackage{makecell}
\usepackage{multirow}   % \multirow in ablation tables
\usepackage{graphicx} 
\usepackage[T1]{fontenc}
\usepackage[utf8]{inputenc}  
\usepackage{dblfloatfix}
\usepackage{siunitx}
\usepackage{enumitem}
\usepackage{hyperref}
\usepackage{placeins}

\newif\ifDEBUG
\DEBUGfalse
\DEBUGtrue

\ifDEBUG   
    \newcommand{\BC}[1]{\textcolor{purple}{[BSHC: #1]}} % Purple
    \newcommand{\YZ}[1]{\textcolor{red}{[YZ: #1]}}     
    \newcommand{\SK}[1]{\textcolor{olive}{[SK: #1]}} 
\else

    \newcommand{\BC}[1]{}
    \newcommand{\YZ}[1]{} 
    \newcommand{\SK}[1]{} 
\fi

\usepackage{cleveref}

\title{ICLAD: In-Context Learning with Comparison-Guidance for Audio Deepfake Detection}
\author{Benjamin Chou$^{\dagger}$ \\
  Purdue University, USA \\
  \texttt{chou150@purdue.edu} \\ \And
  Yi Zhu \\
 Reality Defender Inc.,  USA \\
  \texttt{yi.zhu@inrs.ca} \\ \And 
   Surya Koppisetti \\
 Reality Defender Inc., USA \\
  \texttt{surya@realitydefender.ai} \\ }

\begin{document}
\maketitle

\renewcommand{\thefootnote}{\fnsymbol{footnote}}
\footnotetext[2]{Work done during internship at Reality Defender Inc.} 
\renewcommand{\thefootnote}{\arabic{footnote}}

\begin{abstract}
Audio deepfakes pose a significant security threat, yet current state-of-the-art (SOTA) detection systems do not generalize well to realistic in-the-wild deepfakes. We introduce a novel \textbf{I}n-\textbf{C}ontext \textbf{L}earning paradigm with comparison-guidance for \textbf{A}udio \textbf{D}eepfake detection (\textbf{ICLAD}). The framework enables the use of audio language models (ALMs) for training-free generalization to unseen deepfakes and provides textual rationales on the detection outcome. At the core of ICLAD is a pairwise comparative reasoning strategy that guides the ALM to discover and filter hallucinations and deepfake-irrelevant acoustic attributes. The ALM works alongside a specialized deepfake detector, whereby a routing mechanism feeds out-of-distribution samples to the ALM. On in-the-wild datasets, ICLAD improves macro F1 over the specialized detector, with up to $2\times$ relative improvement. Further analysis demonstrates the flexibility of ICLAD and its potential for deployment on recent open-source ALMs.
% \footnote{Internship work done at Reality Defender Inc.}
\end{abstract}

\section{Introduction}

% The proliferation of generative models has resulted in publicly available tools that can synthesize speech closely mimicking a specific person \cite{le_voicebox_2023}.
% Text-to-speech (TTS) and voice conversion (VC) systems can generate audio deepfakes from mere seconds of authentic speech.
% This technology presents significant societal risks, including the spread of misinformation, eroded trust in voice-based identification, and the forgery of legal evidence \cite{pender_ai_2023,cox_how_2023}.
% Consequently, reliable audio deepfake detection (ADD) has become a critical area of research.

Generative models can now synthesize highly convincing audio deepfakes from a few seconds of human speech~\cite{le_voicebox_2023}, posing significant societal risks of misinformation and identity forgery~\cite{pender_ai_2023,cox_how_2023}. Audio Deepfake Detection (ADD) has hence become a critical area of research. State-of-the-art (SOTA) ADD models typically rely on fine-tuning large self-supervised models on deepfakes generated from scripted speech collected in a studio~\cite{xiao_xlsr-mamba_2025,chen_rawbmamba_2024,truong_temporal-channel_2024,tak_automatic_2022, muller_does_2024}, such as the ASVspoof datasets~\cite{wang_asvspoof_2020,liu_asvspoof_2023,wang_asvspoof_2024}. In contrast to speech recorded "in the wild", scripted studio speech data are collected under controlled and constrained conditions, making it free from real-world extrinsic variations like background noise, room acoustics, and compression artifacts. Furthermore, being non-spontaneous, scripted speech lacks natural disfluencies, such as filled pauses, repetitions, and false starts common in conversation~\cite{nagrani_voxceleb_2017, shriberg_spontaneous_2005, mclaren_speakers_2016}. As a result, performance saturation has been observed on scripted studio deepfakes, accompanied by severe degradation when models are tested on realistic in-the-wild deepfakes~\cite{ge2025post, zhu2025auddtaudiounifieddeepfake}. This generalization gap significantly limits the practical usage of existing detectors, as deepfakes used in real attacks inherently carry in-the-wild characteristics.

% As shown in our preliminary benchmarks (\Cref{tab:macro_f1}), SOTA detectors show a significant performance degradation when evaluated on unscripted, in-the-wild speech ~\cite{muller_does_2024}.
A common way to mitigate this generalization issue is to retrain detectors on expanded training data. However, repetitive supervised fine-tuning is both expensive and not sustainable as new deepfake generation methods continue to emerge rapidly. More importantly, real-world speech data are hard to obtain due to privacy concerns, and their specific patterns differ drastically across use cases (\textit{e.g.}, noisy, monotonic telephony speech from call centers versus highly expressive voice from social media platforms). Hence, a generalizable deepfake detector must be able to adapt quickly and effectively to new conditions, without requiring extensive, costly domain-specific retraining. 

To bridge this gap, we propose \textbf{I}n-\textbf{C}ontext \textbf{L}earning with comparison-guidance for \textbf{A}udio \textbf{D}eepfake detection (\textbf{ICLAD}), a training-free deepfake detection method that generalizes to unseen in-the-wild deepfakes with textual explanations on the classification outcome. The core idea of ICLAD lies in a Pairwise Comparative Reasoning (PCR) strategy, where the ALM is initially prompted to provide real and fake evidence simultaneously from the selected examples, without access to the ground-truth label. The ground-truth is then exposed to the model, enabling an iterative self-discovery process where the ALM identifies and filters deepfake-irrelevant attributes and inherent hallucinations. We further complement the ALM with a specialized deepfake detector to focus on subtle deepfake cues that may be overlooked by the ALM. We found that ICLAD yields higher macro F1 scores than specialized detectors on realistic in-the-wild deepfake datasets, while providing rich textual explanations to support its decision.

% Pretrained on vast and diverse speech corpora, LALMs possess a more generalized understanding of in-the-wild speech characteristics, such as XXX. Furthermore, they exhibit an emergent capability for in-context learning (ICL), which allows adaptation to new tasks from a few examples provided in a prompt, without requiring parameter updates~\cite{liu_what_2021, olsson_iinn-dcuocntitoenxthleeaadrsning_nodate,min_metaicl_2022}. 
% ICL offers a potential solution to the limitations of existing methods by enabling training-free, on-the-fly adaptation to new speech types, thereby mitigating the need for frequent retraining and the use of large amount of real-world speech data for fine-tuning.

\noindent \textbf{Our main contributions are:}
\begin{enumerate}[partopsep=0pt,topsep=0pt,parsep=0pt,leftmargin=*]
\item We introduce ICLAD, built on an audio language model, demonstrating the first successful adaptation of in-context learning (ICL) for training-free detection of audio deepfakes.
\item We design a novel \textit{pairwise comparative reasoning} strategy to guide the ALMs to identify and filter out hallucinations as well as deepfake-irrelevant acoustic attributes.
\item We show that ICLAD improves over the specialized detector on unseen in-the-wild deepfake datasets, while providing textual rationales for ALM decisions.
\end{enumerate}

% https://arxiv.org/pdf/2506.21090

\section{Related Work}

\subsection{Specialized Audio Deepfake Detectors}
The dominant paradigm of ADD has been centered on training classifiers in a fully-supervised manner \cite{yi_audio_2023}.
SOTA systems typically employ self-supervised learning (SSL) encoders, such as Wav2Vec2 \cite{tak_automatic_2022} or WavLM \cite{combei_wavlm_2024}, as frontend feature extractors, with a classification backend to map high-dimensional representations to a binary decision.
While these models achieve near-perfect performance on scripted studio deepfake datasets (ASVspoof 2019, 2021)~\cite{wang_asvspoof_2020,yamagishi_asvspoof_2021}, they fail to generalize to unseen attacks and in-the-wild speech \cite{ muller_does_2024}.
% This performance collapse is a well-documented challenge, often attributed to models learning spurious, dataset-specific artifacts, producing sharp drops in performance when data distributions shift~\cite{muller_does_2024, wang_asvspoof_2025}.
While methods such as data augmentation \cite{tak2021rawboost,tak_automatic_2022}, frontend finetuning \cite{9747768, wang_asvspoof_2024}, and extracting more robust features \cite{zhu_slim_2024, zhu_characterizing_2023} have been proposed to mitigate this issue, they often come at a drastic increase in training cost and still show large discrepancies between performance achieved with scripted studio deepfakes and in-the-wild deepfakes~\cite{muller_does_2024}.
This fundamental generalization gap underscores the need for a more adaptive paradigm that can handle novel threats without requiring constant retraining.

\subsection{Audio Language Models for Audio Deepfake Detection}
The recent advent of ALMs~\cite{kong_audio_2024,ghosh_audio_2025, kimiteam_kimi-audio_2025,goel_audio_2025, xu_qwen25-omni_2025} has introduced a new approach for ADD. Unlike specialized detectors, ALMs are pre-trained on a larger corpus of vast, diverse multimodal data, giving them a more general understanding of real-world speech patterns. However, current audio deepfake detection systems remain dominated by bespoke detectors, with no evidence that off-the-shelf ALMs can be deployed without task-specific adaptation~\cite{gu2025allm4add}. This gap stems largely from a data mismatch: ALMs are trained overwhelmingly on authentic speech (\textit{e.g.}, Common Voice~\cite{ardila_common_2020}) with limited exposure to deepfakes. To bridge this gap, recent work has focused on adapting ALMs through Supervised Fine-Tuning (SFT), often by reformulating deepfake detection as an Audio Question Answering (AQA) task (\textit{e.g.}, ``Is this audio fake or real?'')~\cite{gu2025allm4add}.
While improved performance has been observed on scripted studio deepfakes, this SFT approach still inherits the limitations of costly finetuning and fails to generalize to in-the-wild data~\cite{gu2025allm4add}.

% This presents two significant practical challenges that our work aims to address: (1) it perpetuates the costly ``arms race,'' requiring constant model updates as new attacks emerge, and (2) it is often infeasible in commercial settings where retraining on sensitive or proprietary customer data is prohibited.
% Recent work on adapting self-supervised speech models still relies on analysing internal representations or fine-tuning large encoders to suit downstream objectives~\cite{zhu_slim_2024, xiao_xlsr-mamba_2025,  chen_wavlm_2022}.
% Our work diverges from these training-dependent paradigms by leveraging the training-free adaptation capabilities of in-context learning.

\subsection{In-Context Learning for Adapting to Unseen Data}

ICL is an emergent capability of large-scale models to perform a new task based on a few examples provided in the prompt, without any parameter updates \cite{liu_what_2021, olsson2022incontextlearninginductionheads,min_metaicl_2022}. 
% Research suggests that ICL is not task learning in the traditional sense, but a form of sophisticated pattern matching where the model uses the prompt's examples to infer the expected output structure and label distribution~\cite{dewynter2025incontextlearninglearning}.
\begin{figure*}[h]
    \centering
    \includegraphics[width=1\linewidth]{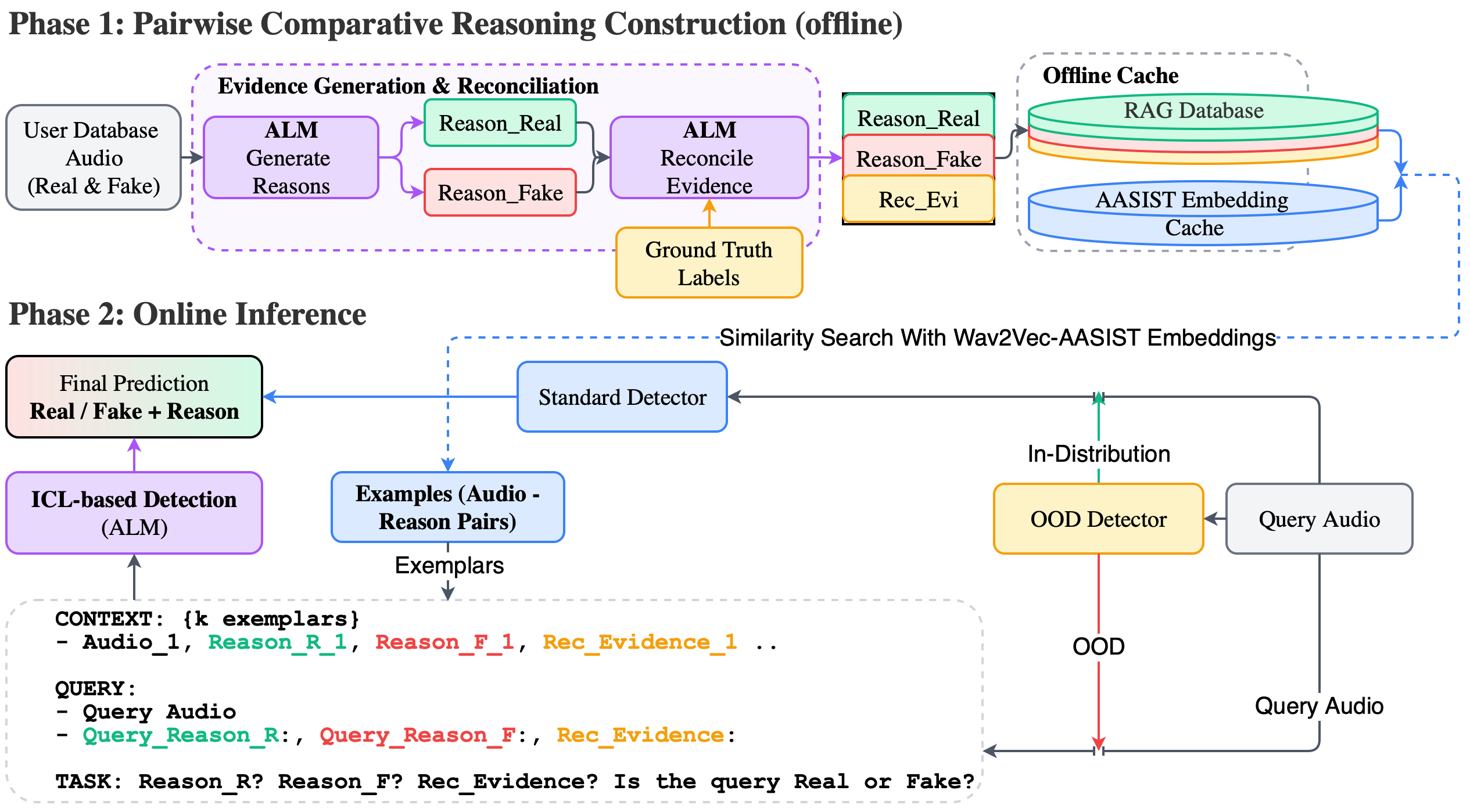}
    % \caption{The ICLAD framework has two phases. In Phase-1 (Section \ref{ref:phase-1}), a Pairwise Comparative Reasoning (PCR) strategy is used to guide an ALM to generate hallucination-aware evidence to support the ground-truth labels.  Phase-2 (Section \ref{sec:phase-2}) retrieves the most acoustically similar examples along with their paired evidence and ground-truth info, and directs the ALM to make a robust final prediction and a reasoning on the query audio.}
    \caption{The ICLAD framework has two phases. In Phase-1 (Section \ref{ref:phase-1}), the ALM is first exposed to a database of labeled audio samples for which it generates evidence supporting both \textit{real} and \textit{fake} classes. Then, in a process we call Pairwise Comparative Reasoning (PCR), the ALM compares this conflicting evidence to produce a reconciled explanation that highlights discriminative attributes consistent with the ground-truth label. During online inference in Phase-2 (Section \ref{sec:phase-2}), an OOD detector is used to route out-of-distribution samples to the ALM. For a given query audio, we retrieve the most acoustically-similar sounding examples from Phase 1, along with their paired reasons, reconciled evidence, and ground-truth label, and employ in-context learning on the ALM to make a robust final prediction and reasoning on the query audio.}
    \label{fig:overview}
\end{figure*}
While ICL allows for rapid, training-free adaptation, the efficacy of ICL is notoriously fragile. Its performance is highly sensitive to the choice of exemplars, and it tends to learn superficial correlations rather than the underlying logic of a task \cite{dewynter2025incontextlearninglearning, chen_relation_2024}. This brittleness leads to two critical challenges: first, ICL performs poorly on tasks requiring reasoning, and second, its performance degrades on Out-of-Distribution (OOD) data \cite{dewynter2025incontextlearninglearning}. Adapting ALMs from their pre-training on bona fide speech to deepfake detection is a clear example of such an OOD challenge. In our proposed framework, ICLAD, this generalization issue is addressed by a new PCR strategy that guides the ALM to self-discover deepfake-relevant attributes by leveraging its general audio understanding capabilities, thus achieving robust generalization.

\section{ICLAD}
\label{sec:method}
% Our methodology is designed to address the key requirements of a modern audio deepfake detection system: generalization to unseen acoustic conditions, robustness against model-specific artifacts, and practical deployability.
% A significant challenge with many contemporary detectors is their failure to generalize to ``in-the-wild'' speech, which represents a substantial distribution shift from the scripted datasets they are trained on \cite{muller_does_2024}.
% This raises a critical question: do these detectors genuinely understand the fundamental traces of deepfakes, or are they merely modeling dataset-specific artifacts?
% Furthermore, while many studies focus on the Equal Error Rate (EER), a threshold-independent metric, we observe that models can be over-optimized for EER and perform poorly in practice where a fixed binary threshold is required (see \Cref{tab:macro_f1}).

\Cref{fig:overview} provides an overview of ICLAD, a comparative reasoning framework designed to achieve training-free generalization for in-the-wild deepfakes while yielding rich textual explanations. ICLAD entails two phases. In phase-1, the PCR strategy is used to guide an ALM to generate hallucination-aware evidence from a diverse sample pool. Phase-2 retrieves the most acoustically similar examples with their paired evidence and ground-truth from the phase-1 database, concatenates them into the ICL prompt, and directs the ALM to make a robust final decision on the query audio. The following sections detail the workflow and components of both phases.

% The offline reasoning process including the pairwise comparative reasoning strategy and Retrieval-augmented generation (RAG) database is detailed in Section.~\ref{sec:phase-1}. The online inference phase is detailed in Section.~\ref{sec:phase-2}.

% This section details our iterative development process. We begin by establishing our core learning paradigm in the ``Audio In-Context Learning'' section.
% We then describe our main contribution, a systematic exploration of prompt engineering to elicit robust analytical reasoning in the ``Prompt Engineering for Robust Reasoning'' section.
% Subsequently, we detail our data-driven selection of an optimal example retrieval mechanism in the ``Example Retrieval'' section.
% Finally, we integrate these components into a practical, hybrid framework that leverages an Out-of-Distribution (OOD) detector to combine the strengths of our adaptive model with specialized detectors in the ``Hybrid Detection Framework'' section.

\subsection{Phase-1: Offline Reasoning}
\label{ref:phase-1}

\begin{figure}[h]
    \centering
    \includegraphics[width=1\linewidth]%{figures/ICL_strategy-2.png}
    {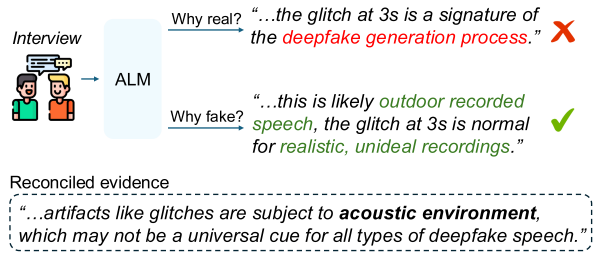}
    \caption{ALMs can cite the same attribute (e.g., a glitch during an interview) as both the sign of a \textit{real} and \textit{fake} speech. Our PCR strategy addresses the issue by forcing a comparison between \textit{real} and \textit{fake} evidence to obtain the \textit{reconciled} evidence, which helps the ALM discover discriminative attributes consistent across datasets.
    }
    % \caption{Visualization of pair-wise comparative reasoning (PCR). Two ALM prompts produce explanations for ``real'' and ``fake'' evidence. Both can hallucinate and cite the same unreliable cues (breath, smooth F0). By comparing the two, PCR filters out unreliable evidence.}
    \label{fig:icl-gap}
\end{figure}
\subsubsection{Motivation for Comparative Reasoning}
We first observed that in a zero-shot setting, ALMs exhibit a strong bias that defaults consistently to a single class prediction (either ``real'' or ``fake''). This demonstrates that ALMs cannot be used as off-the-shelf deepfake detection tools, motivating the use of ICL. Following conventions~\cite{liu_what_2021, NEURIPS2023_398ae57e}, we initially tested a simple \textsc{[Audio]-[Label]} format, interleaving audio tokens and corresponding labels in the prompt before the final query. However, this approach yielded performance no better than random chance. This failure of simple ICL suggests that ALMs could not independently infer the complex acoustic attributes required for deepfake detection, thus motivating our exploration of more sophisticated reasoning.

During our exploration of different ICL strategies, we observed that ALMs can generate contradictory explanations based on ambiguous acoustic cues (\Cref{fig:icl-gap}). For example, when prompted to provide evidence for both authenticity and forgery of the same audio clip, the might cite the presence of a glitch as evidence of the deepfake generation process, while simultaneously using the same cue as evidence of a real speech. We therefore design the PCR strategy to force a paired real and fake evidence comparison. This strategy compels the ALM to reconcile these contradictory rationales with the ground-truth, enabling it to explicitly pinpoint and filter ambiguous cues, while discovering the discriminative deepfake-related attributes.

\subsubsection{Pairwise Comparative Reasoning}
\textbf{Initial Evidence Generation.} For each audio sample $\mathbf{A}_i$ with its corresponding label $\mathbf{L}_i \in \{\text{real}, \text{fake}\}$, the ALM is initially prompted without access to the ground-truth label. The prompt is designed to compel the model to simultaneously generate two sets of textual explanations, namely the real evidence $\mathbf{R}_{real,i}$ and the fake evidence $\mathbf{R}_{fake,i}$. This pairwise evidence generation process forces the ALM to consider both sides of $\mathbf{A}_i$.\\
\textbf{Evidence Reconciliation.} To mitigate the evidence contradiction observed in the initial evidence generation step, we introduce a reconciliation step to discover and filter out ambiguous acoustic cues. For the audio sample $\mathbf{A}_i$, the ALM is prompted with $\mathbf{R}_{real,i}$, $\mathbf{R}_{fake,i}$, along with the ground-truth $\mathbf{L}_i$ to generate a reconciled evidence $\mathbf{R}_{reconciled,i}$. Intuitively, the ALM is tasked to review and reconcile its previously generated (potentially contradictory) descriptions based on the ground-truth. The goal of reconciliation is two-fold. First, we aim to identify and discount acoustic attributes that are not indicative of the true label, while being present. Second, to filter out hallucinated attributes that do not exist in the audio.\\
\textbf{Offline Cache.} The generated evidence $\mathbf{R}_{real,i}$, $\mathbf{R}_{fake,i}$, and $\mathbf{R}_{reconciled,i}$ for all training samples are stored in an offline cache, which serves as a Retrieval-Augmented Generation (RAG)~\cite{lewis_2020} database of high-quality, hallucination-aware explanations. To facilitate fast retrieval during inference time, for each audio sample in the database, we extracted embeddings using a pretrained specialized deepfake detector Wav2Vec2-AASIST~\cite{tak_automatic_2022}, resulting in an associated AASIST embedding cache.

\subsection{Phase-2: Online Inference}
\label{sec:phase-2}
\subsubsection{Example Retrieval}
\label{sec: Retrieval}
Upon receiving a query audio $\mathbf{A}_q$, we first extract the Wav2Vec2-AASIST embeddings, which are used to query the embedding cache constructed in phase-1. This retrieval-augmented process identifies the \textit{K} most acoustically similar exemplar entries from the cache, where each entry includes the audio, label, and evidence information $(\mathbf{A}_K, \mathbf{L}_K, \mathbf{R}_{real,K}, \mathbf{R}_{fake,K}, \mathbf{R}_{reconciled,K})$. To optimize the selection of examples, we experimented with different embedding choices and found Wav2Vec2-AASIST with the best detection results. The comparison is detailed in \Cref{rag}.

\subsubsection{Dynamic routing}
\label{sec: OOD-detection}
While ALM demonstrates strong capabilities in general audio understanding, specialized deepfake detectors may excel at capturing fine-grained deepfake acoustic artifacts. To leverage the complementarity between the two types of models, the query audio $\mathbf{A}_q$ is passed through an Out-of-Distribution detector. This detector assesses which detector $\mathbf{A}_q$ should be routed to. If $\mathbf{A}_q$ is classified as in-distribution (ID) (i.e., similar to studio speech/deepfakes), it is routed to a specialized deepfake detector to obtain a final decision. In this case, ALM is not involved. If $\mathbf{A}_q$ is classified as OOD (i.e., in-the-wild sample), it is sent to the ALM for further processing. For the OOD detector, we employ a standard k-Nearest Neighbor (k-NN) approach~\cite{bukhsh_out--distribution_2023}, implemented with the \textsc{faiss} library. \\
\textbf{Inference.}
For OOD samples, we retrieve \textit{K} examples, each in the format of $(\mathbf{A}_K, \mathbf{L}_K, \mathbf{R}_{real,K}, \mathbf{R}_{fake,K}, \mathbf{R}_{reconciled,K})$, then embed into the ICL prompt. Similar to the reasoning process in phase-1, the ALM is asked to provide real, fake, and reconciled evidence and provide a binary decision (i.e., real or fake).
% ---------------------------------------------

\section{Experimental Setup}

\subsection{Datasets}
We evaluate on five datasets: ASVspoof 2021 (21DF) and MLAAD-v3 (studio-scripted), and ITW, SpoofCeleb, and DFEval 2024 (in-the-wild). These cover 126,348 clips across 42 languages. See Table~\ref{tab:dataset_details_singlecol} and App.~\ref{app:data} for detailed split sizes, languages, and licenses.
% For calibration only, we use ASVspoof2019--LA train/dev; its eval split remains held out.
\subsection{Evaluation Protocol}
\textbf{Data.} We evaluate on \textbf{126{,}348} audio clips across the five corpora. (ASVspoof2021: \textbf{29,738}, MLAAD: \textbf{35,000}, ITW: \textbf{31,280}, SpoofCeleb: \textbf{18,226}, Deepfake-Eval-2024: \textbf{12,104}). For datasets without train/test splits, we subsample the ICL set disjointly from the test set. All audio files are truncated to 4\,s to match the input prepossessing pipeline of the baseline. \\
% \textbf{Ablations.} We run ablations on five-datasets, totaling at \textbf{6{,}965} clips.
\textbf{Metrics.} While Equal Error Rate (EER) has been the dominant performance metric for ADD~\cite{yamagishi_asvspoof_2021, wang_asvspoof_2025}, its reliance on continuous scores (\textit{e.g.}, raw logits) makes it unsuitable for evaluating our framework, where the ALM outputs a binary decision. Furthermore, since the calculation of EER does not require a binarized threshold, a low EER value may not directly translate into high accuracy (see \Cref{fig:logit_dist_appendix}). To better reflect performance in a practical deployment scenario, where a hard classification is required and class imbalance is often present, we report macro F1-score and accuracy as our primary metrics.  It should be noted that a fixed binarization threshold of 0.5 is used for calculating macro F1 and accuracy, rather than relying on the dataset-specific EER threshold. This approach accurately mimics real-world deployment conditions where optimal thresholds cannot be pre-calculated. \\
\textbf{Hardware.} Experiments were conducted on an NVIDIA A100 40 GB GPU.
\subsection{ICLAD Setup}
\textbf{ALM choice.} While ICLAD can be applied to any ALMs, we use Gemini-2.5 Flash as it demonstrates significantly stronger audio understanding capabilities on multiple benchmarks~\cite{kong2025audioflamingosoundcottechnical, geminiteam2025geminifamilyhighlycapable}. We further performed ablations with other open-source ALMs, such as Audio Flamingo 3 (AF3) in ~\Cref{open-source}.\\
\textbf{ICL hyperparameters.} We select samples using embedding similarity as the retrieval mechanism from a RAG database consisting of 500 samples subsampled from 19DF(19DF)~\cite{wang_asvspoof_2020} and 500 from the target set's training split. We select 10 examples as the context for inferring each query audio, with $5$ real and $5$ fake.\\
\textbf{OOD detector.} Our OOD kNN router uses hyperparameters from the original paper~\cite{bukhsh_out--distribution_2023}: ($k{=}5$, $Threshold{=}95\%$).

\subsection{Baseline}
For our comparative analysis, we selected Wav2Vec2-AASIST~\cite{tak_automatic_2022} as the baseline deepfake detector. This choice was motivated by its demonstrated superior performance and stronger generalization capability when evaluated on in-the-wild speech compared to the five other specialized detectors shown in \Cref{tab:macro_f1_appendix}.

\begin{figure}[t]
    \centering
    \includegraphics[width=1\linewidth]{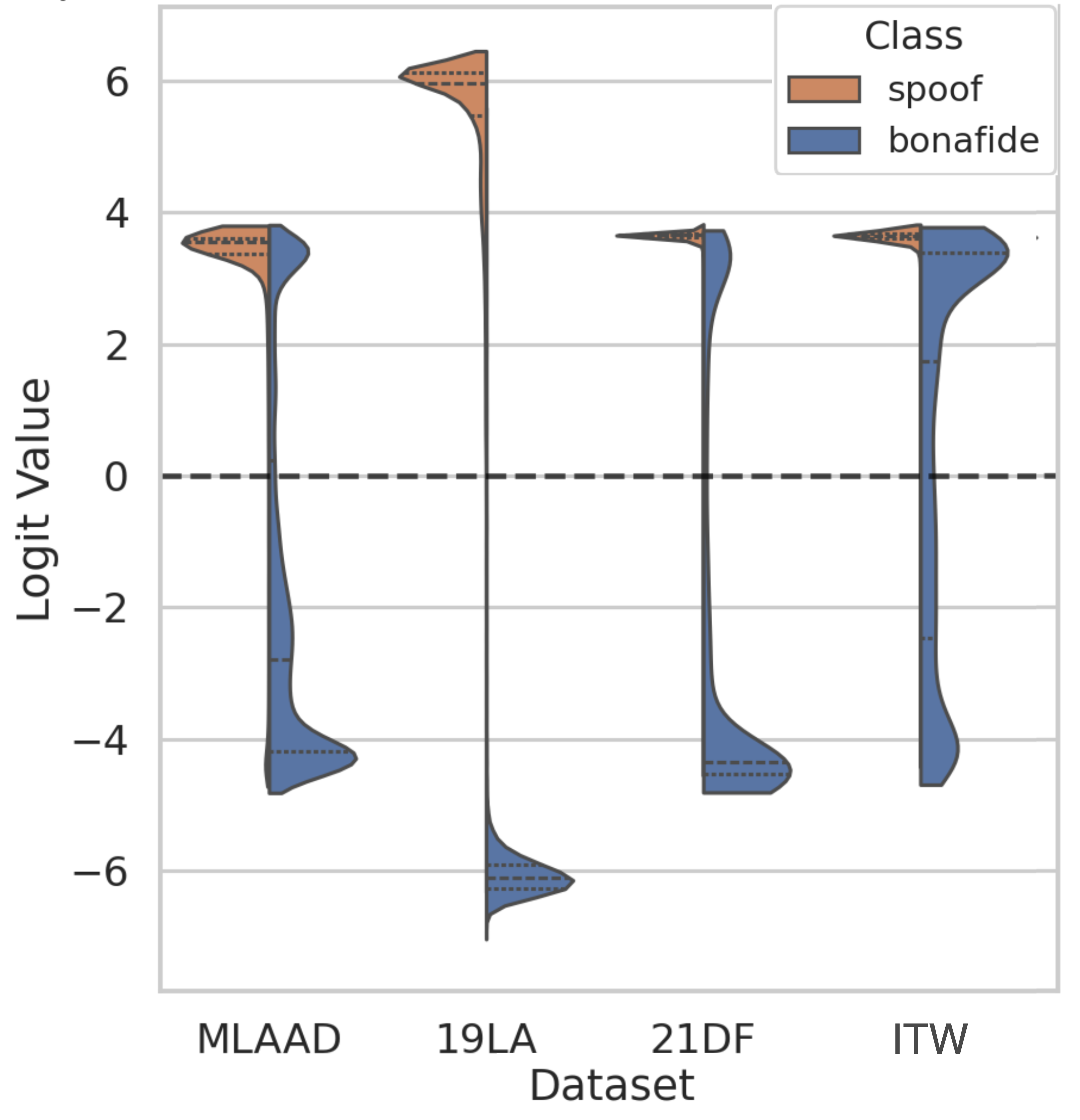}
    \caption{Logit distributions of Wav2Vec2-AASIST on ID (ASVspoof 2021) vs. OOD (ITW, SpoofCeleb) datasets. Overlap between classes is significantly more common on OOD data.}
    \label{fig:logit_dist_appendix}
\end{figure}

\begin{table}[h!]
\centering
\caption{Macro-F1 scores of SOTA detectors. Wav2Vec2-AASIST shows the best overall generalization.}
\label{tab:macro_f1_appendix}
\setlength{\tabcolsep}{3pt}
\begin{tabular}{@{}lccc@{}}
\toprule
\textbf{Model} & \textbf{21DF} & \textbf{ITW} & \textbf{MLAAD} \\
\midrule
XLSR Mamba       & \textbf{0.936} & 0.523 & 0.558 \\
WavLM ASP        & 0.786 & 0.474 & 0.398 \\
XLSR Conformer   & 0.924 & 0.644 & 0.716 \\
RawBMamba        & 0.738 & 0.444 & 0.695 \\
Wav2Vec2-AASIST   & 0.925 & \textbf{0.681} & \textbf{0.733} \\
\bottomrule
\end{tabular}
\end{table}

\noindent \textbf{Logit distributions.} To further understand the generalization gap of specialized detectors, we analyzed the output logit distributions of the Wav2Vec2-AASIST baseline across different datasets (see \Cref{fig:logit_dist_appendix}). On ID data like 19DF, the distributions for real and fake classes are well-separated. However, on in-the-wild datasets, the distributions significantly overlap, leading to poor classification performance and explaining the drop in Macro F1 scores. This highlights the issue with using EER as the primary metric, because the optimal threshold varies between datasets, and a good threshold for one dataset can potentially lead to very poor performance on another.

\section{Results and Analysis}
% We evaluated \textbf{ICLAD} on a set of deepfake detection datasets.
% This section presents the findings, beginning with an overall performance comparison, followed by ablations of the main components.

\subsection{Performance Comparison}
\label{full}
% We compared ICLAD against the baseline detector on subsets of about 20,000 samples per dataset.

\textbf{Generalization to in-the-wild deepfakes.} \Cref{tab:performance_comparison} shows the comparison between ICLAD and the SOTA specialized deepfake detector. The baseline detector retains an expected advantage on its matched training domains (MLAAD and ASVspoof 2021) due to its specialized fine-tuning. In sharp contrast, the generalization of ICLAD is evident across the other three in-the-wild datasets (i.e., ITW, SpoofCeleb, and DFEval 2024), where it consistently outperforms the baseline. The most significant discrepancy occurs on SpoofCeleb, where ICLAD's macro F1 score of 0.665 represents nearly a $2\times$ improvement over the baseline's 0.334. \\
\begin{table}[b]
\centering
\caption{Performance of ICLAD against a specialized detector. Best Macro F1 ($\uparrow$) per dataset is bolded (see \Cref{app:stats} for statistical analysis results).}
\label{tab:performance_comparison}
\setlength{\tabcolsep}{4pt}
\begin{tabular}{@{}lcccc@{}}
\toprule
& \multicolumn{2}{c}{\textbf{Baseline}} & \multicolumn{2}{c}{\textbf{Ours}} \\ 
\cmidrule(lr){2-3} \cmidrule(lr){4-5}
\textbf{Dataset} & \textbf{Acc.} & \textbf{F1} & \textbf{Acc.} & \textbf{F1} \\
\midrule
21DF & 0.868 & \textbf{0.866} & 0.825 & 0.822 \\
MLAAD & 0.800 & \textbf{0.798} & 0.593 & 0.593 \\
ITW & 0.691 & 0.674 & 0.778 & \textbf{0.777} \\
SpoofCeleb & 0.501 & 0.334 & 0.668 & \textbf{0.665} \\ 
DFEval 2024 & 0.500 & 0.367 & 0.550 & \textbf{0.550} \\ 
\bottomrule
\end{tabular}
\end{table}
\begin{table}[h]
\centering
\caption{Correct and incorrect classifications made by ICLAD. GT: Ground-truth. Pred: ICLAD's output.}
\label{tab:qualitative_examples}
\setlength{\tabcolsep}{4pt}
\begin{tabular}{@{}lp{5.75cm}@{}}
\toprule
\multicolumn{2}{@{}l}{\textbf{GT: Fake, Pred: Fake}} \\
\midrule
\textbf{Reasoning} &  Speech flow is overly smooth and consistent, with no micro-hesitations (“um”, “uh”), stutters, or self-corrections. Pauses are timed with machine-like precision, lacking natural irregularities. \\
\midrule
\multicolumn{2}{@{}l}{\textbf{GT: Real, Pred: Real}} \\
\midrule
\textbf{Reasoning} & Natural inhalations occur before phrases (\textit{e.g.}, “conducts”, “uh insane”), and spontaneous fillers (“uh”) are audible. \\
\midrule
\multicolumn{2}{@{}l}{\textbf{GT: Real, Pred: Fake}} \\
\midrule
\textbf{Reasoning} & Although breathing is present, the broader range of physiological noises — lip smacks, tongue clicks, swallows — is missing. \\
\midrule
\multicolumn{2}{@{}l}{\textbf{GT: Fake, Pred: Real}} \\
\midrule
\textbf{Reasoning} &  A clear, naturally timed inhalation at the very start, along with additional breath cues, strongly suggests human physiology. \\
\bottomrule
\end{tabular}
\end{table}

\noindent \textbf{Explainability.} Unlike specialized detectors that output only a score, our model produces textual reasoning alongside its decision, enabling qualitative inspection of the detection process. As illustrated by examples in \Cref{tab:qualitative_examples}, these rationales often focus on synthesis artifacts or physiological markers. At the same time, rationale quality is not guaranteed for every sample; we therefore report hallucination-focused listening-test results for the prompting strategies (\Cref{tab:simple_strategy_eval,tab:listening_test_reason_distribution,app:listening-test}).

% \noindent \textbf{Explainability.} A key advantage of ICLAD is its inherent explainability. Unlike specialized detectors that output only a score, our model generates explicit, rich textual reasoning alongside its decision, enabling detailed qualitative analysis of the detection process. As illustrated by examples in \Cref{tab:qualitative_examples}, the model correctly identifies and prioritizes specific synthesis artifacts or natural physiological markers. Analyzing the failure cases reveals that errors often stem from the model being misled by recording artifacts, providing a clear and actionable path for future improvement.

\subsection{Comparison between ICL strategies}
\textbf{Setup.} We run the ablations on five-datasets, totaling at \textbf{6{,}965} clips. We compare the \textit{pairwise comparative} strategy with two other baseline strategies, namely a \textit{simple} prompting strategy and a \textit{knowledge-guided} prompting strategy. The former introduced a reasoning step to create an $\textsc{(audio | reason | label)}$ structure. The descriptions of the in-context examples were generated offline by prompting the model with the audio and its ground-truth label, compelling it to create a justification. The latter prompts the ALM to analyze a predefined set of acoustic attributes that are deemed discriminative by humans, including intonation and emotion, speech quality and audio artifacts, biological signs, and natural pacing and hesitations. \\

\begin{table}%[h]
\centering
\caption{Prompt ablation results across datasets. Best Macro F1 (\(\uparrow\)) per dataset and on average is in bold.}
\label{tab:prompt-ablation}
\setlength{\tabcolsep}{2pt}
\begin{tabular}{llccc}
\toprule
\textbf{Dataset} & \textbf{Metric} & \textbf{Simple} & \textbf{Explicit} & \textit{PCR} \\
\midrule
\multirow{2}{*}{21DF} & Accuracy & 0.8251 & 0.8267 & 0.8442 \\
& Macro F1 & 0.8210 & 0.8231 & \textbf{0.8422} \\
\midrule
\multirow{2}{*}{MLAAD} & Accuracy & 0.6355 & 0.5551 & 0.6111 \\
& Macro F1 & \textbf{0.6395} & 0.5808 & 0.6110 \\
\midrule
\multirow{2}{*}{\mbox{ITW}} & Accuracy & 0.7844 & 0.7359 & 0.8071 \\
& Macro F1 & 0.8022 & 0.7204 & \textbf{0.8045} \\
\midrule
\multirow{2}{*}{SpoofCeleb} & Accuracy & 0.6213 & 0.5976 & 0.6527 \\
& Macro F1 & 0.6097 & 0.5662 & \textbf{0.6511} \\
\midrule
\multirow{2}{*}{DFEval 2024} & Accuracy & 0.5661 & 0.5489 & 0.5411 \\
& Macro F1 & 0.5554 & \textbf{0.5834} & 0.5410 \\
\midrule[\heavyrulewidth]
\multirow{2}{*}{\textbf{Average}} & Accuracy & 0.6865 & 0.6528 & 0.6917 \\
& Macro F1 & 0.6856 & 0.6548 & \textbf{0.6905} \\
\bottomrule
\end{tabular}
\end{table}

\noindent \textbf{Quantitative results.} \Cref{tab:prompt-ablation} presents the quantitative ablation results, showing that on average the PCR strategy outperforms both the simple and knowledge-guided strategies. The knowledge-guided strategy exhibits the largest performance variance, an expected outcome given its strong inductive bias. This bias is beneficial when its predefined attributes align with a dataset's distribution, but it becomes detrimental in other settings.

However, the superiority of PCR is not uniform. The limitations are most apparent on datasets where real and fake audio share ambiguous or overlapping cues. For example, in MLAAD, the scripted real class contains fewer physiological markers than its fake counterpart. In these challenging scenarios, PCR's reconciled evidence marks most extracted cues as unreliable. Consequently, the model defaults to its zero-shot bias toward predicting audios as real and leads to random chance performance. This reveals a fundamental trade-off: while PCR is an effective technique to filter hallucinated evidence, this process can be overly aggressive, inadvertently removing the discriminative cues required for a reliable in-context analysis. \\

\noindent\textbf{Qualitative analysis.} Since our method forces the ALM to self-explore acoustic attributes associated with deepfake labels, a common issue that we observed is hallucination, where non-existent attributes are generated in the textual explanation to justify the ground-truth. While hallucination may not necessarily lead to a decrease in accuracy, it will result in incorrect textual explanations. In our qualitative analysis, 22 human annotators with audio analysis experience were recruited to identify whether an audio recording is paired with hallucinated ALM explanation. \Cref{tab:simple_strategy_eval} first provides an overview of different hallucination categories observed using the simple prompting strategy. Across 120 generated explanations, 18.3\% has hallucinations, where prosody and naturalness related attributes are shown as the leading category. In contrast, only 10.0\% of the explanation generated were identified with hallucinations using the PCR strategy. We further provide per-sample annotation result in \Cref{app:listening-test}.

\begin{table}[h]
\centering
\caption{Distribution of hallucination categories. Majority voting results indicate that the leading category is naturalness (37.4\%).}
\label{tab:listening_test_reason_distribution}
\setlength{\tabcolsep}{3pt}
\small
\begin{tabular}{@{}p{0.46\columnwidth}cc@{}}
\toprule
\textbf{Category of reason} & \textbf{Count} & \textbf{Proportion (\%)} \\
\midrule
Prosody/Naturalness (pitch, intonation, pacing) & 46 & 37.40 \\
Other (semantic, uncategorized) & 36 & 29.27 \\
Physiological signals (breathing, throat) & 20 & 16.26 \\
Acoustics/artifacts (noise, distortion, clipping) & 19 & 15.45 \\
Unseen language & 2 & 1.63 \\
\midrule
\textbf{Total} & \textbf{123} & \textbf{100.00} \\
\bottomrule
\end{tabular}
\end{table}

With knowledge-guided prompting, we notice two distinct problems. First, without explicit guidance to treat the attributes separately, the ALM tends to describe different acoustic aspects with high correlation. For example, a single perceived cue in one category would cause it to invent conforming evidence in the others, resulting in a high hallucination rate of approximately 50\%.

% \noindent\textbf{Qualitative analysis.} With simple prompting, the model is observed to suffer from hallucinations, where plausible but non-existent attributes are generated to justify the ground-truth, thereby failing to learn genuine discriminative cues. A manual evaluation of 120 generated explanations revealed that 18.3\% contained such hallucinations (see \Cref{tab:simple_strategy_eval}). With knowledge-guided prompting, we notice two distinct problems. First, without explicit guidance to treat the attributes separately, the ALM tends to describe different acoustic aspects with high correlation. For example, a single perceived cue in one category would cause it to invent conforming evidence in the others, resulting in a high hallucination rate of approximately 50\%. 
% While adding a direct instruction to ``Analyze each of the following attributes independently'' successfully reduced this rate to around 10\%, we encountered a more fundamental issue: \textbf{expert bias}.
Secondly, we find that introducing human knowledge can severely bias ALM when testing on certain datasets. For example, the model learns to associate the absence of biological signs (\textit{e.g.}, breaths or pauses) with fake audio. This heuristic failed on datasets like MLAAD, where the real speech is often professionally recorded and scripted, naturally containing fewer biological markers. 
\begin{table}[h]
\centering
\caption{Human evaluation of explanations generated using the \textit{Simple} strategy. N=120.}
\label{tab:simple_strategy_eval}
\begin{tabular}{@{}lcc@{}}
\toprule
\textbf{Rating} & \textbf{Count} & \textbf{Percentage} \\
\midrule
High & 64 & 53.3\% \\
Medium & 36 & 30.0\% \\
Low & 20 & 16.7\% \\
\midrule
\textbf{Hallucinations} & 22 & 18.3\% \\
\bottomrule
\end{tabular}
\end{table}

% By forcing the model to focus only on attributes that we, as experts, deemed important, we inadvertently limited its ability to discover the most truly discriminative cues. This led to poor generalization and worse performance (see Table \ref{tab:prompt-ablation}). For example, the model learned to heavily associate the absence of biological signs (like breaths or pauses) with fake audio. This heuristic failed on datasets like MLAAD, where the real speech is often professionally recorded and scripted, naturally containing fewer of these ``biological'' markers. The model, biased by the attribute list, would incorrectly classify this clean, real audio as fake. This demonstrated that a truly robust system needed to seek out and learn the reliability of different acoustic cues, directly motivating our final \textit{Pair-wise comparative} strategy.

\subsection{Retrieval Embedding Ablation}
\label{rag}
\begin{table*}[!t]
\centering
\small
\renewcommand{\arraystretch}{0.95}
\caption{RAG ablation results across datasets. `Detector' shows the performance of our non-ICL baseline: Wav2Vec2-AASIST. Best Macro F1 (\(\uparrow\)) per dataset is in bold, second best is underlined.}
\label{tab:rag-ablation}
\setlength{\tabcolsep}{4pt}
\begin{tabular}{llccccc}
\toprule
\textbf{Dataset} & \textbf{Metric} & \textbf{Detector} & \textbf{Wav2Vec2} & \textbf{AASIST} & \textbf{Text} & \textbf{AASIST+Text} \\
\midrule
\multirow{2}{*}{21DF} & Accuracy & 0.9162 & 0.8415 & 0.8442 & 0.7740 & 0.8463 \\
& Macro F1 & \textbf{0.9148} & 0.8394 & 0.8422 & 0.7656 & \underline{0.8443} \\
\midrule
\multirow{2}{*}{MLAAD} & Accuracy & 0.8025 & 0.6041 & 0.6111 & 0.5199 & 0.6182 \\
& Macro F1 & \textbf{0.7999} & 0.6030 & 0.6110 & 0.5029 & \underline{0.6163} \\
\midrule
\multirow{2}{*}{\mbox{ITW}} & Accuracy & 0.6479 & 0.7552 & 0.8071 & 0.7403 & 0.7881 \\
& Macro F1 & 0.5998 & 0.7483 & \textbf{0.8045} & 0.7237 & \underline{0.7850} \\
\midrule
\multirow{2}{*}{SpoofCeleb} & Accuracy & 0.5008 & 0.6497 & 0.6527 & 0.6441 & 0.6326 \\
& Macro F1 & 0.3396 & 0.6462 & \textbf{0.6511} & 0.6472 & \underline{0.6287} \\
\midrule
\multirow{2}{*}{DFEval 2024} & Accuracy & 0.4990 & 0.5441 & 0.5411 & 0.4988 & 0.5170 \\
& Macro F1 & 0.3654 & 0.5441 & \textbf{0.5435} & 0.5160 & \underline{0.5170} \\
\midrule[\heavyrulewidth]
\multirow{2}{*}{\textbf{Average}} & Accuracy & 0.6733 & 0.6789 & 0.6912 & 0.6354 & 0.6804 \\
& Macro F1 & 0.6039 & 0.6762 & \textbf{0.6905} & 0.6311 & \underline{0.6783} \\
\bottomrule
\end{tabular}
\end{table*}

\textbf{Setup.} We run the ablations on five-datasets, totaling at \textbf{6{,}965} clips. We conducted a systematic comparison of four different embedding choices to identify the optimal one for example retrieval. These embeddings include Wav2Vec2-XLSR~\cite{baevski_wav2vec2_2020}, Wav2Vec2-AASIST~\cite{tak_automatic_2022}, Qwen3-0.5B text embeddings computed from the evidence~\cite{yang2025qwen3technicalreport}, and a combination of the audio and text embeddings. With the former three, we adopted cosine similarity for finding the most similar examples; with the audio+text embeddings, Maximal Marginal Relevance (MMR) was used to find examples with maximized audio similarity for relevance and minimized text similarity for diversity consideration.\\

\noindent \textbf{Results.} \Cref{tab:rag-ablation} compares the performance achieved using different embedding strategies for RAG, benchmarked against the specialized baseline \textbf{`Detector'}. Results confirm that Wav2Vec2-AASIST embeddings lead to the best performance on average, surpassing the baseline detector by 8.66\%. This suggests that task-specific embeddings are more effective for finding acoustically meaningful examples than relying on general-purpose audio or text representations. In contrast, using text embeddings leads to consistently poorer performance. This is potentially caused by the high similarity in the textual description of real and fake classes, which may cause the PCR mechanism to overly aggressively filter attributes as unreliable. We attempted to exploit this phenomenon by combining Wav2Vec2-AASIST embedding similarity with text dissimilarity for retrieval, which led to marginally better performance on ID datasets (21DF: +0.21\%; MLAAD: +0.53\%). However, the approach degraded performance on in-the-wild datasets, indicating that a simple text dissimilarity is insufficient. Further investigation is required to effectively integrate textual semantics into the retrieval process.

\FloatBarrier
\subsection{Importance of Dynamic Routing}
\label{Main Results}
\begin{table}[!ht]
\centering
\caption{OOD ablation. Best Macro F1 (\(\uparrow\)) per dataset is in bold, second best is underlined.}
\label{tab:ood-ablation}
\setlength{\tabcolsep}{2pt}
\begin{tabular}{llrr}
\toprule
\textbf{Dataset} & \textbf{Strategy} & \textbf{Accuracy} & \textbf{Macro F1} \\
\midrule
\multirow{3}{*}{21DF} & PCR & 0.6460 & 0.6456 \\
& Baseline & 0.9162 & \textbf{0.9148} \\
& OOD & 0.8442 & \underline{0.8422} \\
\midrule
\multirow{3}{*}{MLAAD} & PCR & 0.5265 & 0.5120 \\
& Baseline & 0.8025 & \textbf{0.7999} \\
& OOD & 0.6111 & \underline{0.6110} \\
\midrule
\multirow{3}{*}{ITW} & PCR & 0.6427 & \underline{0.6424} \\
& Baseline & 0.6479 & 0.5998 \\
& OOD & 0.8071 & \textbf{0.8045} \\
\midrule
\multirow{3}{*}{SpoofCeleb} & PCR & 0.5593 & \underline{0.5577} \\
& Baseline & 0.5008 & 0.3396 \\
& OOD & 0.6527 & \textbf{0.6511} \\
\midrule
\multirow{3}{*}{DFEval 2024} & PCR & 0.5374 & \underline{0.5281} \\
& Baseline & 0.4980 & 0.3648 \\
& OOD & 0.5411 & \textbf{0.5410} \\
\bottomrule
\end{tabular}
\end{table}
\Cref{tab:ood-ablation} compares the performance obtained with and without dynamic routing. Results clearly reveal a performance trade-off between the two core components: the specialized baseline maintains its superiority on ID datasets (MLAAD and 21DF), while the ICL framework demonstrates superior generalization on the diverse in-the-wild datasets (ITW, SpoofCeleb, and DFEval 2024).

The dynamic routing is designed to capitalize on this trade-off. It effectively routes ID samples to the baseline detector; for 21DF and MLAAD, it routes the majority of samples to the specialized detector, leading to significant macro F1 increases on 21DF (19.6\%) and MLAAD (+9.9\%). Furthermore, even on in-the-wild datasets, the OOD detector successfully identifies a small but critical subset of ID data where the specialized detector remains most accurate. This dynamic routing mechanism helps ICLAD exceed the performance of both the ICL and the specialized detector alone, yielding better macro F1s on all three in-the-wild datasets.

% Specifically, vs PCR, we see gains of 16.21, 9.34, and 1.29 F1 percentage points on the progressively more recent ITW (2022), SpoofCeleb (2024), and DFEval 2024 (2025) datasets, respectively. This clear trend of decreasing gains reflects the growing disparity between modern in-the-wild deepfakes and the original ASVspoof datasets.
\subsection{Open-Source ALM Evaluation}
\label{af3}
\label{open-source}
\begin{table}[!ht]
\centering
\caption{Comparison between Gemini-2.5 Flash and Audio Flamingo 3 (AF3) using the \textit{simple} ICL strategy. *For this test, AF3 is prompted using explanations pre-generated by Gemini.}
\label{tab:opensource-comparison}
\setlength{\tabcolsep}{5pt}
\begin{tabular}{@{}llcc@{}}
\toprule
\textbf{Dataset} & \textbf{Model} & \textbf{Accuracy} & \textbf{Macro F1} \\
\midrule
\multirow{2}{*}{21DF} & AF3* & \textbf{0.6772} & \textbf{0.6507} \\
& Gemini & 0.6192 & 0.6086 \\
\midrule
\multirow{2}{*}{ITW} & AF3* & \textbf{0.7890} & \textbf{0.7885} \\
& Gemini & 0.6951 & 0.6909 \\
\bottomrule
\end{tabular}
\end{table}

We evaluate AF3 (7B parameters) as an open-source alternative to Gemini-2.5 Flash. However, AF3's instruction-following capabilities were severely limited, consistently outputting only a binary label without intermediate reasoning (See \Cref{app:af3_failures}). To enable comparison, we bypassed phase-1 and directly provided AF3 with the explanations generated by Gemini as examples and used the \textit{simple} prompting strategy for ICL (\Cref{tab:opensource-comparison}).

Surprisingly, with pre-generated explanations from Gemini, AF3 obtains higher accuracy than Gemini-2.5 Flash. This suggests AF3 may possess good audio understanding capability but lacks the necessary thinking capability, likely due to its training objective being focused on direct output. This finding suggests a path forward for a fully open-source and more accurate version of ICLAD.

% This suggests that fine-tuning AF3 for CoT may enable a fully open-source and more accurate version of ICLAD. However, for now, this hybrid approach allows us to incur a one-time cost to produce the intermediate explanations with a closed-source ALM. The inference step can then be done locally via an open-source model like AF3, without sacrificing performance.

\section{Conclusion}
We introduce ICLAD, a deepfake detection paradigm that leverages in-context learning capabilities of ALMs to improve generalization to in-the-wild audio deepfakes. The ALM is guided with a Pairwise Comparative Reasoning (PCR) strategy, in order to generate evidence on discriminative per-class artifacts that are consistent across diverse audio samples. In our experiments, ICLAD improves over the specialized detector on three in-the-wild datasets, while the specialized detector remains stronger on scripted in-distribution datasets. ICLAD also produces textual rationales, and our listening-test analysis indicates lower hallucination rates for PCR than for simple prompting. The proposed framework is flexible to operate with recent open-source ALMs (\textit{e.g.}, Audio Flamingo 3), supporting its practical deployment potential.

% We introduce ICLAD, a deepfake detection paradigm that leverages in-context learning capabilities of ALMs to achieve generalization to in-the-wild audio deepfakes. The ALM is guided with a Pairwise Comparative Reasoning (PCR) strategy, in order to generate evidence on discriminative per-class artifacts that are consistent across diverse audio samples. ICLAD is seen to outperform SOTA deepfake detectors across three in-the-wild datasets, while also providing rich textual explanations on the decisions. The proposed framework is flexible to operate with recent open-source ALMs (\textit{e.g.}, Audio Flamingo 3), confirming viability for real-world deployments.

% We presented a framework for audio deepfake detection that uses in-context learning with large audio language models.
% By applying a PCR prompting strategy, which makes the model reason over conflicting evidence, detection of in-the-wild data improves.
% By combining this with an OOD router and a pretrained detector, the hybrid system reaches the best results on out-of-distribution datasets.
% This shows the potential of ICL for generalization in audio deepfake detection and suggests new directions for future work.

% \section*{Acknowledgments}

\section*{Limitations}
% \BC{This does not count toward page limit}
A current limitation is ICLAD’s performance degradation on scripted-speech datasets like MLAAD. Our results suggest that while ALM reasoning captures high-level inconsistencies, it cannot yet fully replace specialized detectors for low-level acoustic artifacts in studio settings. A further limitation of our work is its reliance on proprietary Gemini-2.5 Flash for its offline evidence generation and reconciliation phase. This dependency on a proprietary model was necessary due to the instruction-following capabilities required for the comparative reasoning strategy, which we found lacking in currently available open-source ALMs. However, our results demonstrate that recent open-source ALMs, such as Audio Flamingo 3, can obtain better performance when using Gemini-generated cues. With further improvements to instruction-following capabilities, open-source ALMs can be prompted with the proposed reasoning strategy, which enables a completely self-contained, highly performant version of the ICLAD framework. 
\clearpage
\bibliography{references}

\clearpage
\appendix
\section{Appendix}
\subsection{Potential Risks}
% \BC{Yi, pls fill in here}
The reliance of our framework on ALMs may introduce risks related to algorithmic bias. For example, biases that might be present in the ALMs' training data (\textit{e.g.}, gender, accent, language) may unintentionally affect the detection process and certain user groups. However, the preliminary results obtained with the open-source AF3 model suggest that a fully transparent system capable of bias discovery and auditing is feasible for future implementation.

\subsection{Listening Test Results}
\label{app:listening-test}
To investigate the reliability of ALM generated textual explanations and how they align with human perception, we conducted a systematic listening test to quantify model hallucination rate using the proposed Pairwise Comparative Reasoning (PCR). We evaluated the ALM's PCR outputs using 50 randomly selected audio samples (10 per benchmark dataset). We recruited 22 human annotators with verified experience in audio and speech processing. Each annotator was tasked with validating the PCR's reasoning against the actual audio characteristics. To ensure high-quality feedback, each annotator focused on a specific attribute: speech content, prosody/naturalness, room acoustics, physiological cues, or background noise/artifacts. Annotators chose between three labels: (i) no hallucination, (ii) hallucination (requiring a specific reason), or (iii) unsure. A hallucination was strictly defined as reasoning describing an audible event demonstrably absent from the sample. We employed an overlapping design, assigning 20 samples per person to ensure each sample received multiple expert labels. Following rigorous quality checks, 8 annotators were excluded due to missing data or failure to follow the hallucination definition, leaving a robust set of expert-validated labels.

\subsubsection{Quantitative Analysis}
We applied majority voting across annotations to determine the final hallucination status for each sample. The detailed distribution of these human labels per sample is provided in \textbf{Table~\ref{tab:listening_test_human_distribution}}.

Overall, the results indicate that the PCR strategy significantly mitigates model errors. Only 10\% of ALM responses using PCR contained hallucinations, a marked improvement over the 18\% hallucination rate observed with a simple prompting strategy (as noted in Table 4 of the main manuscript). This reduction demonstrates that the comparative framework effectively grounds the ALM's reasoning in actual acoustic evidence.

To understand the nature of the remaining errors, we categorized the reasons provided by annotators for "hallucinated" labels using keyword matching. As shown in \textbf{Table~\ref{tab:listening_test_reason_distribution}}, the primary category for disagreement is \textit{Naturalness} (37.40\%). Analysis of these cases reveals a consistent contradiction: ALMs frequently label scripted speech with a steady, robotic pace as "unnatural," regardless of whether the source is real or synthetic.

This specific failure mode explains why the ALM excels on in-the-wild spontaneous speech but struggles with scripted datasets like ASVspoof and MLAAD. These findings validate that while PCR reduces hallucinations, the model's internal bias toward "naturalness" in spontaneous speech remains a target for future alignment.

\begin{table}[t]
\centering
\caption{Human annotation distribution. Each sample received annotations from multiple experts to ensure robust and high-confidence labels.}
\label{tab:listening_test_human_distribution}
\setlength{\tabcolsep}{3pt}
% \scriptsize
\small
\begin{tabular}{@{}cccc@{}}
\toprule
\textbf{Sample} & \textbf{\% Hall.} & \textbf{\% Not Hall.} & \textbf{\% Unsure} \\
\midrule
9 & 69.2 & 30.8 & 0.0 \\
11 & 57.1 & 42.9 & 0.0 \\
3 & 50.0 & 37.5 & 12.5 \\
28 & 50.0 & 50.0 & 0.0 \\
5 & 33.3 & 66.7 & 0.0 \\
1 & 28.6 & 71.4 & 0.0 \\
26 & 28.6 & 71.4 & 0.0 \\
8 & 28.6 & 71.4 & 0.0 \\
15 & 27.3 & 72.7 & 0.0 \\
13 & 25.0 & 75.0 & 0.0 \\
18 & 25.0 & 50.0 & 25.0 \\
2 & 25.0 & 75.0 & 0.0 \\
16 & 25.0 & 62.5 & 12.5 \\
30 & 25.0 & 75.0 & 0.0 \\
23 & 22.2 & 77.8 & 0.0 \\
14 & 22.2 & 77.8 & 0.0 \\
21 & 20.0 & 80.0 & 0.0 \\
27 & 20.0 & 80.0 & 0.0 \\
22 & 20.0 & 60.0 & 20.0 \\
20 & 18.2 & 81.8 & 0.0 \\
17 & 10.0 & 80.0 & 10.0 \\
12 & 10.0 & 90.0 & 0.0 \\
7 & 9.1 & 90.9 & 0.0 \\
24 & 7.7 & 84.6 & 7.7 \\
19 & 0.0 & 100.0 & 0.0 \\
10 & 0.0 & 100.0 & 0.0 \\
25 & 0.0 & 100.0 & 0.0 \\
6 & 0.0 & 83.3 & 16.7 \\
4 & 0.0 & 66.7 & 33.3 \\
29 & 0.0 & 100.0 & 0.0 \\
\bottomrule
\end{tabular}
\end{table}

\subsection{Statistical Significance Testing}
\label{app:stats}
\begin{table}[h]
\centering
\caption{Paired t-test statistics for the accuracy comparison between the baseline and ICLAD. All results are statistically significant.}
\label{tab:stat_tests_appendix}
\begin{tabular*}{\columnwidth}{@{\extracolsep{\fill}} l r l l @{}}
\toprule
\textbf{Dataset} & \multicolumn{1}{c}{\textbf{$t$-statistic}} & \textbf{$p$-value}  \\
\midrule
 21DF & 8.97  & $< .001$  \\
MLAAD         & 61.11 & $< .001$  \\
ITW  & -7.82 & $< .001$  \\
SpoofCeleb    & -32.95& $< .001$  \\
DFEval 24   & -7.98 & $< .001$  \\
\bottomrule
\end{tabular*}
\end{table}

To verify the observed performance differences between the baseline and our Gemini ICL framework, we performed paired t-tests on the results. As detailed in \Cref{tab:stat_tests_appendix}, the results are statistically significant ($p < .001$) across all five datasets.

\subsection{Datasets}
\label{app:data}
\begin{table}[h]
\centering

\setlength{\tabcolsep}{1pt}
\caption{
    Dataset details. No model training was performed. ICL examples were drawn from a database of 500 samples from the target's train split and 500 from 19DF.
    \textbf{Licenses}: ODC-By (Open Data Commons Attribution); Apache-2.0 (Apache License 2.0); CC BY 4.0 (Creative Commons Attribution 4.0); CC BY-SA 4.0 (Creative Commons Attribution-ShareAlike 4.0).
}
\label{tab:dataset_details_singlecol}
\begin{tabular}{@{}lcccl@{}} % Changed lrrrl to ccccc
\toprule
& \multicolumn{2}{c}{\textbf{Test Size}} & \makecell{\textbf{RAG}\\\textbf{(Train)}} & \\
\cmidrule(lr){2-3} \cmidrule(lr){4-4}
\textbf{Dataset} & \textbf{Main} & \textbf{Abl.} & & \textbf{License} \\
\midrule
19DF & --- & --- & 500 & ODC-By \\
21DF & 29,738 & 1,394 & 500 & ODC-By \\
MLAAD & 35,000 & 2,210 & 500 & Apache-2.0 \\
ITW & 31,280 & 1,160 & 500 & Apache-2.0 \\
SpoofCeleb & 18,226 & 1,200 & 500 & CC BY 4.0 \\
DFEval 2024 & 12,104 & 1,000 & 500 & CC BY-SA 4.0 \\
\bottomrule
\end{tabular}
\end{table}
We evaluate our framework on five deepfake datasets representing two distinct conditions: scripted studio speech and challenging in-the-wild audio. For scripted studio datasets we use \textbf{ASVspoof 2021 (21DF)}~\cite{yamagishi_asvspoof_2021} and \textbf{MLAAD}~\cite{muller_mlaad_2025}. 21DF contains \textbf{English} read-speech from the VCTK corpus. We use a subset of MLAAD that covers the eight languages present in both spoofed and real audio (German, Polish, English, French, Italian, Spanish, Russian, and Ukrainian). The in-the-wild datasets feature spontaneous speech from public figures. \textbf{In-the-Wild (ITW)}~\cite{muller_does_2024} contains audio of \textbf{58 politicians and celebrities} collected from social networks and video platforms. \textbf{SpoofCeleb}~\cite{jung_spoofceleb_2025} is built upon the VoxCeleb1 dataset, featuring voices from \textbf{1,251 celebrities}. The \textbf{DFEval 2024}~\cite{chandra_deepfake-eval-2024_2025} benchmark contains content from 88 websites and \textbf{42 languages}, with its audio subset being 78.7\% \textbf{English}. \textbf{ASVSpoof 2019 (19DF)}~\cite{wang_asvspoof_2020} is used exclusively to supplement the RAG database. \Cref{tab:dataset_details_singlecol} provides a detailed breakdown of data splits and licenses. We follow the intended use of all the datasets.

\subsection{Instruction-Following Failures in Audio Flamingo 3}
\label{app:af3_failures}
As noted in \Cref{af3}, Audio Flamingo 3 (AF3) exhibited significant limitations in following complex instructions for structured data generation. \Cref{tab:af3_failures_revised} provides several examples of these failure modes. In each case, AF3 was prompted to provide a reasoned analysis within a specific JSON schema. However, the model frequently produced outputs that violated the prompt's constraints, such as omitting required rationales or echoing schema placeholders verbatim.

We attempted to mitigate these issues using generation enforcement libraries like \texttt{lm-format-enforcer}\cite{lmformatenforcer_2023} and \texttt{xgrammar}\cite{dong2024xgrammar}. While these tools forced AF3 to produce syntactically valid JSON, they could not prevent the model from oftentimes generating semantically illogical content.
\begin{table}[h!]
\centering
\caption{Examples of instruction-following and logical failures in AF3.}
\label{tab:af3_failures_revised}
\setlength{\tabcolsep}{5pt}
\begin{tabularx}{\columnwidth}{@{} X @{}}
\toprule
% --- Omitted Rationale ---
\multicolumn{1}{@{}l}{\textbf{Omitted Rationale}} \\
\cmidrule(r){1-1}
The model returns a schema but leaves the required analytical fields empty. \newline
\textit{Example:} \texttt{\{"Reconciled\_Evidence": ""\}} \\

\addlinespace[10pt] % Add space between entries

% --- Echoed Placeholders ---
\multicolumn{1}{@{}l}{\textbf{Echoed Placeholders}} \\
\cmidrule(r){1-1}
Instead of generating new content, the model copies placeholder text from the prompt verbatim. \newline
\textit{Example:} \texttt{\{"Final\_Answer": "real | fake"\}} \\

\addlinespace[10pt]

% --- Format Violation ---
\multicolumn{1}{@{}l}{\textbf{Format Violation}} \\
\cmidrule(r){1-1}
The model ignores a JSON-only instruction and outputs a free-form prose sentence. \newline
\textit{Example:} ``The audio clip is real'' \\

\addlinespace[10pt]

% --- Illogical Content ---
\multicolumn{1}{@{}l}{\textbf{Illogical Content}} \\
\cmidrule(r){1-1}
    \begin{minipage}[t]{\linewidth}
    \vspace{0pt}
    The model produces semantically nonsensical and syntactically valid output. Examples:
    \begin{itemize}[nosep, leftmargin=*, after=\strut]
        \item Returning the audio transcription\newline
\textit{Example:} ``Because men groping in the Arctic darkness had found a yellow metal''
        \item Filling competing hypothesis fields with the same irrelevant text.
        \item A non-meaningful response \newline
\textit{Example:} ``The audio clip is a recording of a human voice''
    \end{itemize}
    \end{minipage} \\
\bottomrule
\end{tabularx}
\end{table}

\FloatBarrier
\subsection{LLM Use in Manuscript Preparation}
\label{appendix-llm}
We used OpenAI's GPT-5 (via ChatGPT) to help with wording clarity and grammar edits.
All scientific claims, experimental design, data analysis, and conclusions remain the responsibility of the authors.

\end{document}